\documentstyle[aas2pp4]{article}

\tightenlines

\newcommand {\xten}[1]	{\mbox{$\times 10^{#1}$}}
\newcommand {\ArI}	{Ar\,{\sc I}}
\newcommand {\ArIl}[1]  {\ArI~$\lambda$#1}

\newcommand {\NAr} {N_{\rm Ar}}
\newcommand {\nAr} {n_{\rm Ar}}

\lefthead{Parker et al.}
\righthead{The Spectroscopic Detectability of Argon in the Lunar Atmosphere}

\begin{document}

\title{The Spectroscopic Detectability of Argon in the Lunar Atmosphere}

\author{Joel~Wm.~Parker and S.~Alan~Stern}
\affil{Department of Space Studies, Southwest Research Institute,
Suite 426, 1050 Walnut Street, Boulder, CO 80302 \\
Electronic mail: joel@boulder.swri.edu, alan@boulder.swri.edu}
\authoraddr{Southwest Research Institute, Suite 426, 1050 Walnut Street,
Boulder, CO 80302}

\author{G. Randall Gladstone}
\affil{Department of Space Science, Southwest Research Institute,
6220 Culebra Road, San Antonio, TX 78238 \\
Electronic mail: randy@whistler.space.swri.edu}

\and

\author{J. Michael Shull\altaffilmark{1}}
\affil{Center for Astronomy and Space Physics, Department of Astrophysical and
Planetary Science, University of Colorado, Campus Box 389, Boulder, CO 80309 \\
Electronic mail: mshull@casa.colorado.edu}
\altaffiltext{1}{Also at JILA, University of Colorado and
 National Institute of Standards and Technology}

\begin{abstract}

Direct measurements of the abundance of argon in the lunar atmosphere were made
in 1973 by instruments placed on the Moon during the Apollo 17 mission, but the
total daytime abundance is unknown due to instrument saturation effects; thus,
until we are able to return to the Moon for improved direct measurements, we
must use remote sensing to establish the daytime abundance.  In this paper, we
present a complete analysis of the potential for measuring argon in the lunar
atmosphere via emission-line or absorption-line observations.  We come to the
surprising conclusion that the lower limit established by the {\em in situ\/}
lunar argon measurements implies that {\em any\/} absorption-line measurement
of argon in the lower, dayside lunar atmosphere requires analysis in the
optically-thick regime.  In light of this result, we present the results of our
EUVS sounding rocket observations of the lunar occultation of Spica, which
provide a new upper limit on the abundance of argon in the daytime lunar
atmosphere.  We also re-analyze a recently reported weak detection of lunar
atmospheric \ArIl{1048}\ in emission by the {\em ORFEUS\/} satellite, and show
that those data are inconsistent with the emission being due to argon over a
wide range of temperatures (up to at least 2000~K).  This result is primarily
due to our use of a more complete curve of growth analysis, and improved values
for the argon fluorescent emission rates from radiation and solar wind
interactions.  We find that the detection reported by {\em ORFEUS\/} would imply
an argon surface density significantly greater than the total surface density
of the lunar atmosphere for argon accommodated to typical daytime surface
temperatures ($\sim 400$~K), and also is inconsistent with a high-density
transient event.  Therefore, we conclude that the reported argon detection is
untenable.

\end{abstract}

\keywords{atomic data --- line: formation --- radiative transfer --- methods: data analysis --- Moon --- planets and satellites: Moon ---ultraviolet: solar system}

\bigskip

\begin{flushright}
\fbox{To apper in {\em  The Astrophysical Journal Letters\/},
Volume 509, December 1998}
\end{flushright}


\section{Introduction}

The atmosphere of the Moon is a tenuous, surface-boundary exosphere.  The known
neutral constituents of that atmosphere include He, Ar, Rn, Po, Na, and K
(Hoffman et al.~1973; Gorenstein \& Bjorkholm~1973; Potter \& Morgan 1988;
Tyler et al.~1988) with surface number densities that vary with local time of
day and other factors.  The total inventory of the identified lunar atmospheric
neutral species\footnote{Ions are not an important factor in this case as they
are quickly removed by solar wind interaction; see also the detection of O$^+$,
Al$^+$, Si$^+$, and possibly P$^+$ by Mall et al.~(1998).} has a number density
at the surface of $\lesssim 5 \times 10^4$~cm$^{-3}$.  By contrast, cold
cathode gauges placed on the lunar surface during Apollo missions measured
total pressures of the lunar atmosphere corresponding to a total number density
near the surface of $\sim 2 \times 10^{5}$~cm$^{-3}$ at nighttime, and possibly
almost two orders of magnitude higher during the daytime ($\sim 5 \times
10^{6}$ to $1 \times 10^{7}$~cm$^{-3}$), though much of the daytime values
appear to be due to equipment contamination in the landing area (Johnson et
al.~1972).  These number density and total pressure results indicate that most
of the lunar atmosphere remains compositionally unidentified (see Stern [1998]
for a detailed review of the lunar atmosphere).  There are ongoing efforts to
determine and observe the remaining constituents of the lunar atmosphere (e.g.,
Flynn \& Stern 1996; Stern et al.~1997; Mall et al.~1998).

One long-standing possibility for resolving this ``missing mass'' discrepancy
is that argon could comprise a greater fraction of the lunar atmosphere than
existing measurements indicate.  The most direct measurements of the abundance
of argon in the lunar atmosphere were made by the Apollo 17 surface-based mass
spectrometer, LACE\@.  Argon, which is adsorbable on the cold-trapped lunar
surface at night ($T \sim 100$~K), was observed by LACE to follow a diurnal
pattern with a nighttime minimum near 2\xten{2}~cm$^{-3}$, followed by a rapid
increase around sunrise to values as high as 4\xten{4}~cm$^{-3}$ before LACE
became saturated by gas evolving off the warming lunar surface and outgassing
products of nearby Apollo equipment (Hoffman et al.~1973; Hodges \&
Hoffman~1974; Hodges et al.~1974).  Because LACE saturated due to such
contamination shortly after each sunrise, it is not known how far the daytime
column abundance of argon increases above the saturation limit of the
instrument, and it is conceivable that argon provides the bulk of the missing
mass of the daytime lunar atmosphere.

Flynn (1998; hereafter F98) recently reported results of an experiment to
measure the abundance of lunar argon.  That experiment was a search for the
1048~\AA\ and 1067~\AA\ resonance fluorescence emission lines of \ion{Ar}{1}
using the Berkeley spectrograph (Hurwitz et al.~1998) aboard the {\em
ORFEUS-SPAS II\/} satellite, which flew for several days during Shuttle mission
STS-80 in late 1996.  F98 reported a weak ($3\sigma$) detection of the
\ArIl{1048}\ line; the 1067~\AA\ line was not detected, a point we
will discuss later.  F98 analyzed this detection assuming optically-thin
emission, and deduced a surface density of $\nAr=(8\pm3) \times
10^{5}$~cm$^{-3}$.  Because this density is at odds with thermal model
predictions (Hodges et al.~1974), F98 interpreted the result as evidence for a
non-thermal source of argon.

In what follows, we present analyses of \ArI\ absorption line measurements
(such as have been made with our EUVS sounding rocket instrument during a
recent lunar occultation of Spica) and emission line measurements (such as the
{\em ORFEUS\/} observations).  We show that the LACE results imply that {\em
any\/} absorption measurement of argon on the dayside limb of the Moon will be
optically thick.  Similarly, the argon emission line reported by F98 would
correspond to a line-of-sight column density too large to be analyzed in the
optically-thin limit, and in fact would correspond to a surface density of
$\nAr \gtrsim 5\xten{7}$~cm$^{-3}$ in the case of argon at a typical daytime
surface temperature of 400~K.  This number density substantially exceeds the
total lunar atmospheric surface number density (Johnson et al.~1972).  We
further show that the non-detection of the \ArIl{1067}\ line in the {\em
ORFEUS\/} data rules out the possibility of this detection being a real signal
of a fortuitously-observed, high-density transient event, or of a hot component
produced by non-thermal processes.


\section{Analysis \label{sec:disc}}


\subsection{Absorption Measurements}

\begin{figure}[ht]

\plotone{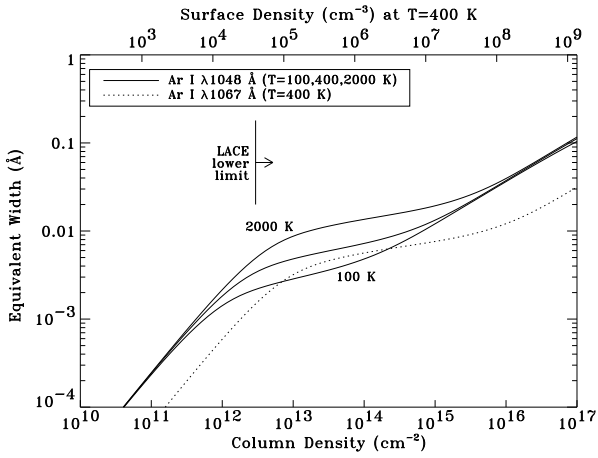}

\caption[]{\sl The line-of-sight column density (bottom axis scale) vs.
equivalent width curve of growth for argon in the lunar atmosphere.  The solid
lines show the behavior of the \ArIl{1048}\ line at three different
temperatures assuming a Voigt profile: $T=100$~K (lower line), $T=400$~K
(middle line), and $T=2000$~K (upper line).  The dashed line shows the curve of
growth for the \ArIl{1067}\ line at a temperature of $T = 400$~K.  The scale on
the top axis of the plot shows the associated surface densities for the case of
$T=400$~K for observations made at a tangential distance from the surface of
one scale height (51~km).  The arrow shows the lower limit of the surface
density of argon, $\nAr > 4 \times 10^4$ cm$^{-3}$, as measured {\em in situ\/}
by the Apollo LACE surface-based mass spectrometer instrument (Hoffman et
al.~1973).  This limit implies that any absorption measurement of argon in the
lower, daytime lunar atmosphere must be calculated in the optically-thick
regime of the curve of growth.
\label{fig:cog} }

\end{figure}

Curves of growth, relating the equivalent width ($W_\lambda$) vs. line-of-sight
column density ($\NAr$), for the \ArIl{1048}\ and $\lambda 1067$ lines for the
case of absorption measurements are shown in Figure~\ref{fig:cog}.  The plot
shows the three well-known regions:  the optically-thin (``linear''), the
doppler (``flat'' or ``logarithmic''), and the damping (``square-root'')
regimes of the curve of growth.  To determine the transition points between
optically-thin and doppler regions, we use the definition of the optical depth
at line center for a single-component Gaussian line profile:

\begin{eqnarray}
\tau_0 & = & \frac{\sqrt{\pi} e^2}{m_e c b} \NAr \lambda f_\lambda \\
       & = & 0.0150 \frac{\NAr \lambda f_\lambda}{b} \label{eq:tau0} ,
\end{eqnarray}

\noindent where $f_\lambda$ is the oscillator strength for the transition at
wavelength $\lambda$ and $b=\sqrt{2kT/m_{\rm Ar}}$ is the doppler velocity
parameter; other symbols have their conventional meanings.  We assume a
Maxwellian velocity distribution at a lunar surface (exobase) temperature of
$T=400$~K (the same temperature as is used by F98), which implies a velocity of
$b = 0.407$~km~s$^{-1}$ for argon atoms.  This temperature is appropriate for a
gas accommodated to the lunar surface daytime equilibrium temperatures.  The
argon oscillator strengths are $f_{1048} = 0.244$ and $f_{1067} = 0.067$,
respectively (Federman et al.~1992); other published values for the oscillator
strengths (e.g., Wiese et al.~1960; Morton~1991; Chan et al.~1992; and other
references in Table~1 of Federman et al.~1992) typically differ by less than
10\% from this adopted value.  The resulting critical column densities
corresponding to the onset of saturation at $\tau_0 = 1$ in the transition from
the optically-thin to the doppler regime are
$\NAr^{\tau=1}(\lambda 1048)=1.1\xten{12}$~cm$^{-2}$ and
$\NAr^{\tau=1}(\lambda 1067)=3.8\xten{12}$~cm$^{-2}$.

To convert column densities (as one observes remotely) into surface densities
(as LACE measured {\em in situ\/}), we use $\nAr = \frac{\NAr}{H \xi}$, where
$H$ is the barometric scale height (for $T=400$~K, the argon scale height at
the surface is $H = 50.9$~km).  The factor $\xi$ is the relationship between
vertical and line-of-sight column densities in cases where the scale height of
the atmosphere is much smaller than the size of the object:  $\xi = \sqrt{2 \pi
d / H}$, where $d$ is the distance of the observation from the lunar center.
At a height of one scale height above the lunar limb, $\xi = 14.5$.  These are
the values that have been used to convert the column densities to the surface
density scale shown on the top axis of Figure~\ref{fig:cog}.

Thus, the critical surface densities where the corresponding line-of-sight
optical depths reach unity are:
$\nAr^{\tau=1}(\lambda 1048)=1.5\xten{4}$~cm$^{-3}$ and
$\nAr^{\tau=1}(\lambda 1067)=5.0\xten{4}$ cm$^{-3}$.
Recall that the argon reported by the LACE mass spectrometer saturated the
instrument at 4\xten{4}~cm$^{-3}$ while increasing just after sunrise.  {\em
Therefore, our analysis implies that any absorption-line measurement of argon
in the lower, daytime atmosphere must use an optically-thick curve of growth
analysis.\/}  This is a significant result that has not been previously
recognized.

In 1995 we used the extreme-ultraviolet spectrograph, EUVS (Slater et al.~1995;
Stern et al.~1996), flown aboard a sounding rocket, to perform such an
absorption-line experiment.  The EUVS consists of a 40~cm diameter Wolter Type
II grazing-incidence telescope feeding a Rowland-circle spectrograph.  On 15
April 1995 we flew EUVS out of White Sands Missile Range to observe the lunar
occultation of the bright B1~V star, Spica ($\alpha$ Vir).  One aspect of this
experiment was its ability to search for lunar argon in absorption, using light
from Spica as the incident beam.  Our EUVS experiment did not detect argon, and
set a $3\sigma$ upper limit of $W_\lambda < 0.043$~\AA, determined by the
background fluctuation in the spectrum around the argon line wavelengths.  For
the expected 400~K atmospheric daytime temperature, this non-detection implies
a line-of-sight density upper limit of $\NAr(\lambda 1048) <
1.3\xten{16}$~cm$^{-2}$ (see Figure~\ref{fig:cog}), and an associated argon
surface density upper limit of $\nAr < 1.8 \xten{8}$~cm$^{-3}$.


\subsection{Emission Measurements}

\begin{figure}[ht]

\plotone{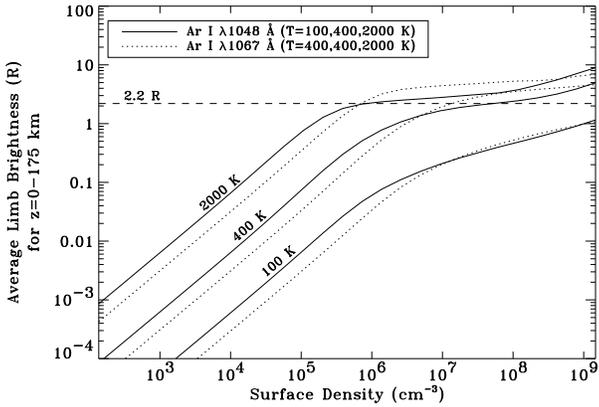}

\caption[]{\sl The surface density vs\@. line brightness curve of growth
for sunlight resonantly scattered by argon in the lunar atmosphere.  The line
brightnesses were calculated by averaging over a range of tangential limb
heights from $z=0$ to 175~km to model the scanning observations made by {\em
ORFEUS\/}.  As in Figure~\protect\ref{fig:cog}, the solid lines show the
behavior of the \ArIl{1048}\ line at three different temperatures: $T=100$~K
(lower line), $T=400$~K (middle line), and $T=2000$~K (upper line).  The dotted
lines show the curves of growth for the \ArIl{1067}\ line at the same
temperatures.  The dashed line labeled ``2.2~R'', indicates the brightness of
the possible emission line near 1048~\AA\ seen in the {\em ORFEUS\/} data.  For
the case of argon accommodated to typical daytime surface temperatures of $\sim
400$~K, the argon surface density implied by the 2.2~R line brightness would be
$\nAr \approx 5.2\xten{7}$~cm$^{-3}$, which substantially exceeds the total
lunar atmospheric number density.  Also, for a wide range of temperatures up to
at least 2000~K, the \ArIl{1067} line (not detected in the {\em ORFEUS\/} data)
would be {\em brighter\/} than the \ArIl{1048} line.
\label{fig:emission} }

\end{figure}

Figure~\ref{fig:emission} shows the curves of growth for emission lines.  The
results are presented in the figure using units that are readily comparable to
the data presented in F98, with line brightness as a function of surface
density (rather than equivalent width as a function of column density as shown
in the curve of growth in Figure~\ref{fig:cog}).  The curves display the
familiar optically-thin/doppler/damping progression of regimes as the surface
density is increased.

To calculate the data for Figure~\ref{fig:emission}, the resonance line
radiative transfer model of Gladstone (1988) was used to simulate the
limb-viewing brightnesses of the two lunar argon lines at the sub-solar
position.  The solar flux at these wavelengths is dominated by the carbon
continuum.  We used flux values of 2.02\xten{7} and
3.20\xten{7}~photons~cm$^{-2}$~s$^{-1}$~\AA$^{-1}$ at 1048 and 1067~\AA,
respectively, based on SOHO/SUMER measurements obtained in 1996 (Wilhelm et
al.~1998).  These fluxes correspond to radiation $g$-factors of 5.0\xten{-8}
and 2.0\xten{-8}~s$^{-1}$, respectively.  We also calculated the effect of
solar wind electron impact based on WIND 3-D plasma instrument (Lin et
al.~1995) data from Nov-Dec 1996, and accepted electron impact excitation cross
sections (Ajello et al~1990).  Our calculations and those of Shemansky
(personal communication, 1998) using these observed flux values show that the
total $g$-factors (radiation + solar wind, as would be appropriate for the
lunar atmosphere during the {\em ORFEUS\/} observations) are about
6.0\xten{-8}~s$^{-1}$ for the \ArIl{1048} line, and 2.6\xten{-8}~s$^{-1}$ for
the \ArIl{1067} line.

F98 used a $g$-factor of 2.2\xten{-7}~s$^{-1}$ for \ArIl{1048}, which was based
on radiation and solar wind flux values that were not appropriate for the
date of the {\em ORFEUS\/} observations.  In particular, the WIND data show
that the actual electron temperature in the solar wind was lower than the
temperature used to calculate the $g$-factor given in F98.  The result is that
our $g$-factor, based on the measured solar radiation and wind fluxes, is 3.7
times lower than that used by F98.

The {\em ORFEUS\/} observations were made using a scanning technique over the
Moon.  The spectrum in F98 showing the line around 1048~\AA\ was obtained
primarily from data within 90~arcsec of the lunar day-side limb; 90~arcsec
corresponds to about 175~km (roughly 3.5 scale heights) at the Moon.  To
calculate the brightness of a line that would be observed in such a scanning
observation, we averaged values from lines of sight with tangential heights
ranging from $z=0$ to $z=175$~km in our model.  Assuming an isothermal
atmosphere at the given temperatures, the model surface density of argon was
varied to obtain the curves of growth for the lines shown in
Figure~\ref{fig:emission}.

F98 gives a purported \ArIl{1048}\ line flux of $F=(1.3 \pm 0.4) \times
10^{-3}$~photons~cm$^{-2}$~s$^{-1}$; the solid angle of the {\em ORFEUS-SPAS
II\/} 20~arcsec aperture is $\Omega = 7.4 \times 10^{-9}$~sr, implying a line
brightness of $B = 4 \pi \frac{F}{\Omega} 10^{-6} = 2.2 \pm 0.7$~Rayleighs.
Our results, shown in Figure~\ref{fig:emission}, suggest that in the case of
$T=400$~K, the surface density of argon that would be required to produce the
observed emission is $\nAr \approx 5.2\xten{7}$~cm$^{-3}$, which is more than a
factor of 60 greater than the density calculated by F98 using the
optically-thin approximation and at the same temperature,  and substantially
exceeds the total lunar atmospheric number density (Johnson et al.~1972).  We
also note that a global surface density of $\nAr = 5.2\xten{7}$~cm$^{-3}$ would
imply a total mass of argon in the lunar atmosphere of $M_{\rm Ar} =
7\xten{9}$~g and a source input rate of $\dot{M}_{\rm Ar} =
6\xten{7}$~g~s$^{-1}$.

Most of this discrepancy between our results and those of F98 for the global
surface density of argon is due to the different $g$-factors used, as discussed
above.  If in our calculations we used the $g$-factor used by F98, we would
find a surface density of $\nAr \approx 9.3\xten{5}$~cm$^{-3}$ (at the
transition between the optically-thin and doppler regimes of the curve of
growth), consistent with the value calculated by F98.  But as we have shown
above, the $g$-factor used by F98 was based on solar radiation and wind flux
values that were not correct for the date of the observations.

Is it possible that the emission line seen in the {\em ORFEUS\/} data
could have been the result of observing a transient event that significantly
enchanced the local density of argon?  Figure~\ref{fig:emission} shows that the
brightness of the \ArIl{1067}\ line, at the implied density
(5.2\xten{7}~cm$^{-3}$) and $T=400$~K, would be significantly {\em brighter\/}
than the \ArIl{1048}\ line.\footnote{At low surface densities (in the optically
thin regime of Figure~\ref{fig:emission}), the brightness of the
\ArIl{1067}\ line is less than the brightness of the \ArIl{1048}\ line by the
ratio of the $g$-factors.  However, as the line cores become optically thick,
the oscillator strength ceases to matter, and the 1067~\AA\ line becomes
brighter due to the larger (factor of 1.6) solar flux.  Since the
1048~\AA\ line has a larger damping constant ($\gamma$) than the
1067~\AA\ line, it enters into the square-root region relatively early, and so
again becomes the brighter emission.} The model results suggest that if the
emission at 1048~\AA\ in the {\em ORFEUS\/} data were due to \ArI, then the
1067~\AA\ line should have been readily detected at a brightness of about
3.2~R.  Yet, no emission line at 1067~\AA\ was detected in the {\em ORFEUS\/}
data, implying that argon was not observed at this density under any
circumstances.

We also considered the possibility of an unknown, non-thermal argon source as
postulated by F98.  Potter \& Morgan (1998) find that sodium in the lunar
atmosphere has a hot component with a typical temperature of 1280~K, and a
maximum of 1736~K.  However, unlike the case for sodium, there is no obvious
mechanism for sufficient non-thermal heating (i.e., sputtering) of noble gases
such as argon (R.~Johnson, personal communication, 1998).  Among our
considerations were the possibility of hot argon vapor production by
micrometeoritic impact on the lunar surface, and argon from the solar wind.  In
the former case, calculations show that the amount of hot argon released into
the atmosphere is an insignificant fraction of the total argon density
(R.~Killen, personal communication, 1998), and in the latter case, the
impacting argon is unlikely to get back into the atmosphere before it has
thermalized.  Still, for completeness, it is useful to address the possibility
of a non-thermal component in case there may be other unknown methods to
produce a significant source of hot argon.  If we assume a supra-thermal
equivalent temperature of 2000~K, the density implied by the \ArIl{1048}\ line
would be 8.4\xten{5}~cm$^{-3}$ (see Figure~\ref{fig:emission}), consistent with
the density calculated by F98.  But even in this extreme case, the
\ArIl{1067}\ line still would be brighter than the \ArIl{1048}\ line, and
should have been detected in the {\em ORFEUS\/} data at a brightness of 2.4~R.
In any case, supra-thermal heating of argon in the lunar atmosphere, if
possible, is probably insignificant, and the argon is most likely to be at the
accommodated temperature of 400~K as discussed earlier.

For these reasons, we conclude that the feature at 1048~\AA\ seen in the F98
data is not due to lunar atmospheric \ion{Ar}{1}\@.  However, our analysis
shows that the best available upper limit for the density of argon in the
daytime lunar atmosphere comes from the non-detection of the \ArIl{1067}\ line
in the {\em ORFEUS\/} spectrum.  As shown in Figure~3 of F98, the approximate
$3\sigma$ limit for the non-detection of a line at $\lambda=1067$~\AA\ is about
1.5\xten{-3}~photons~cm$^{-2}$~s$^{-1}$, corresponding to a brightness of
2.5~R.  According to the $T=400$~K model in our Figure~\ref{fig:emission}, this
implies an upper limit of $\nAr < 2.0 \xten{7}$~cm$^{-3}$, which is
approximately the same as the upper limit for the total density of the lunar
atmosphere from the Apollo measurements (Johnson et al.~1972).


\section{Conclusions}

We find that {\em any\/} abundance measurements based on absorption
observations of argon in the lower, daytime lunar atmosphere must be calculated
in the optically-thick regime of the curve of growth.  The common assumption
that argon in this part of the lunar atmosphere would be optically thin is
invalid.

Also, it appears that the weak emission-line detection at 1048~\AA\ in the F98
{\em ORFEUS\/} data is not due to lunar argon since our analysis shows that: (1)
for argon accommodated to typical daytime surface temperatures of $\sim 400$~K,
the implied argon surface density actually would be significantly larger than
the known density of the entire lunar atmosphere; and (2) at that density, the
\ArIl{1067} line would be brighter than the \ArIl{1048} line, yet no line at
1067~\AA\ is detected in the {\em ORFEUS\/} data, ruling out the possibility
that the emission line is due to a high-density, transient event.  In fact,
even over a wide range of temperatures (up to at least 2000~K) the \ArIl{1067}
line would be brighter than the \ArIl{1048} line, constraining the hot
component of any potential supra-thermal source of argon.  Our results differ
from those of F98 primarily because we applied a more complete curve of growth
analysis, we used improved values for the argon $g$-factors from radiation and
solar wind interactions, and the fact that the implied densities of argon are
not in the optically-thin regime.

The ultimate result of our analysis is that the global density of argon in the
lunar atmosphere is still unknown, and the missing mass mystery remains
unsolved.


\acknowledgements

We thank B.~Flynn for his useful comments on this paper, D.~Shemansky for
discussions and additional $g$-factor calculations, and R. Johnson, R. Killen,
and T. Morgan for other helpful input.  Support for this work was provided by
NASA/Planetary Astronomy grant NAG5-4275 and suborbital rocket grant
NAG5-5006.



\end{document}